\documentstyle[prl,aps]{revtex}
\begin{document}


\vskip 4mm

\centerline{\large \bf Resonant electron transfer between quantum dots}

\vskip 2mm

\centerline{Leonid A. Openov}

\vskip 2mm

\centerline{\it Moscow State Engineering Physics Institute
(Technical University)}
\centerline{\it 115409 Moscow, Russia}
\centerline{E-mail: opn@supercon.mephi.ru}

\vskip 4mm

\begin{quotation}

An interaction of electromagnetic field with a nanostructure composed of two
quantum dots is studied theoretically. An effect of a resonant electron
transfer between the localized low-lying states of quantum dots is predicted.
A necessary condition for such an effect is the existence of an excited bound
state whose energy lies close to the top of the barrier separating the
quantum dots. This effect may be used to realize the reversible quantum logic
gate NOT if the superposition of electron states in different quantum dots is
viewed as the superposition of bits 0 and 1.

\end{quotation}

\vskip 6mm

\centerline{\bf 1. INTRODUCTION}

\vskip 2mm

One way to overcome the limitations of present semiconductor microelectronics
is to reduce the dimensions of electronic devices well below 100 nm size
range. Novel device concepts are based on the use of quantum effects in
nanostructures \cite{Luth}. While a great number of technological problems
still remains to be resolved, there is a considerable experimental and
theoretical activity in the field. Among other things, an interaction of
electromagnetic field with nanostructures is of particular interest since it
results in a variety of phenomena highlighting the wave nature of electrons
(see, e.g., \cite{Krash}).

Grossmann {\it et al.} \cite{Grossmann} have shown that a laser with
appropriate power and frequency can force the electron in a double-well
nanostructure to stay in one of the wells. In this paper we draw attention to
a possibility of an opposite effect, a laser-induced electron transfer
between two quantum dots situated so far from each other that an electron
placed initially in one of the dots may be thought of as localized in that
dot, while having been transferred to the other dot, the electron remains
localized in it after the laser pulse is off.

The paper is organized as follows. We begin with qualitative estimates of
characteristic energies and times of a double-dot nanostructure and
description of an appropriate model. Next we study the temporal evolution of
an electron under the influence of a classical electromagnetic field making
use of the resonant approximation. We demonstrate that the frequency,
amplitude and duration of an electromagnetic pulse may be adjusted in such a
way that an electron will be transferred from the localized lowest-energy
state of one of the quantum dots to the localized lowest-energy state of
another quantum dot with a probability close to unity. We discuss a
possibility of using this effect to realize the quantum logic gate NOT.

\vskip 6mm

\centerline{\bf 2. QUALITATIVE ESTIMATES}

\vskip 2mm

We consider two semiconducting quantum dots, $A$ and $B$, such that each
quantum dot, when isolated, has two size-quantized energy levels in the
conduction band. Let us denote the energies of the lower level
$|\alpha 1\rangle$ and the upper level $|\alpha 2\rangle$ by
$\varepsilon_{\alpha 1}$ and $\varepsilon_{\alpha 2}$ respectively, where
$\alpha=$ $A$ or $B$ is the dot index, and the energies are measured from the
bottom of the conduction band. For the sake of simplicity we assume the
quantum dots to be identical, i.e., the values of
$\varepsilon_{\alpha 1}\equiv\varepsilon_1$ and
$\varepsilon_{\alpha 2}\equiv\varepsilon_2$ do not depend on $\alpha$
($\varepsilon_2 > \varepsilon_1$). The wave functions
$\langle{\bf r}|\alpha 1\rangle$ for the dots $\alpha=$ $A$ and $B$ have the
same functional form but are centered in different regions of coordinate
space, this is also true for the wave functions
$\langle{\bf r}|\alpha 2\rangle$ of the excited states.

If the distance $d$ between the quantum dots and the height $U$ of the energy
barrier separating the dots are reasonably large, the wave functions
$\langle{\bf r}|\alpha 1\rangle$ for $\alpha=$ $A$ and $B$ are strongly
localized in the vicinity of the corresponding quantum dot within the region
of the dot size $a$. Hence, their overlap can be neglected. In other words,
the state $|\alpha 1\rangle$ with the energy $\varepsilon_1$ may be thought
of as doubly degenerate with respect to the dot index $\alpha$, i.e., with
respect to electron location, either in the dot $A$ or in the dot $B$.

It should be stressed that an electron may be considered as localized in one
of the dots in the state $|\alpha i\rangle$ ($i$ = 1 or 2 specifies the
energy level) if we are interested in the processes whose characteristic
times are much shorter than the time $\tau_i$ it takes for an electron to
turn between the states $|Ai\rangle$ and $|Bi\rangle$. The value of $\tau_i$
may be estimated as
\begin{equation}
\tau_i\approx\hbar/V_i,
\label{TAUi}
\end{equation}
where $V_i$ is the energy of electron hopping between the states
$|Ai\rangle$ and $|Bi\rangle$. According to Landau and Lifshitz
\cite{Landau}, in the quasi-classical approximation one has
\begin{equation}
V_i\approx\frac{\hbar}{T_i}\exp\left(-\frac{d}{\hbar}
\sqrt{2m^*(U-\varepsilon_i)}\right),
\label{Vi1}
\end{equation}
where $T_i=\sqrt{2m^*a^2/\varepsilon_i}$ is a period of a classical motion
for an electron with the energy $\varepsilon_i$ in the quantum dot, and $m^*$
is the electron effective mass (for the sake of simplicity, we assume the
values of $m^*$ to be the same in the dots and in the barrier). For the
quantum dot cubic in shape, the ground state energy $\varepsilon_1$ may be
estimated as $\varepsilon_1\approx3\pi^2\hbar^2/2m^*a^2$ provided that
$\varepsilon_1<<U$. Then
$T_i\approx\hbar\pi\sqrt{3/\varepsilon_1\varepsilon_i}$,
and one has from (\ref{Vi1}):
\begin{equation}
V_i\approx\frac{1}{\gamma}\sqrt{\varepsilon_1\varepsilon_i}\exp\left(
-\gamma\frac{d}{a}\sqrt\frac{U-\varepsilon_i}{\varepsilon_1}\right),
\label{Vi2}
\end{equation}
where $\gamma=\pi\sqrt{3}$ is a numerical coefficient.

Taking $U\approx$ 1 eV, $\varepsilon_1\approx$ 0.1 eV, and $d/a\approx$ 3, we
obtain $V_1\sim 10^{-23}$ eV. Hence, a characteristic time it takes for an
electron to tunnel from the ground state of one dot to the ground state of
another dot, $\tau_1\sim 10^{8}$ s, is very long even on a macroscopic scale,
and such a tunneling can be ignored. On the other hand, if the energy
$\varepsilon_2$ of the excited bound state is close to $U$, then the value of
$V_2$ is many orders of magnitude greater than $V_1$. Taking, e.g.,
$U-\varepsilon_2\approx 0.01$ eV, we have $V_2\sim 10^{-3}$ eV and
$\tau_2\sim 10^{-12}$ s. Thus, for a certain set of quantum dots parameters,
the low-lying energy level of the dots can be viewed as degenerate, whereas
the excited level splits into two sublevels with the energies
$\varepsilon_2\pm V_2$. It is important for the following consideration that
the electron wave functions of the resulting excited sublevels are not
localized within a particular dot, but spread over both dots as
$(\langle{\bf r}|A2\rangle \pm \langle{\bf r}|B2\rangle)/\sqrt{2}$.

Of course, our estimates of $V_i$ and $\tau_i$ are rather crude, they
strongly depend on the supposed form of confinement potential and should be
considered as qualitative. However, one can hope that, first, for a dot of an
arbitrary shape it is possible to shift the energy of one of excited states
very close to the continuum part of the energy spectrum by varying, e.g., the
dot size and the doping level, and, second, the distance between two such
quantum dots can be adjusted to satisfy the condition $V_2>>V_1$ for the
energies of electron hopping between the excited states and between the
ground states of the dots respectively (and hence, the condition
$\tau_2<<\tau_1$ for the times of electron switching between those pairs of
states).

It should be pointed out that except for the times $\tau_1$ and $\tau_2$,
there is one more important time scale, the lifetime $\tau^*$ of electron in
the excited state with respect to spontaneous transition to the ground state
at the sacrifice of photon or phonon emission. It has been shown by Nomoto
{\it et al.} \cite{Nomoto} that the value of $\tau^*$ strongly depends on the
dot size and can be as long as 10$^{-6}$ s or even longer, so that one can
suppose $\tau_2 << \tau^* << \tau_1$.

From the above line of reasoning, we set $V_1=0$ (i.e., $\tau_1=\infty$). We
denote $V_2\equiv V$. The diagram of one-electron energy levels is shown
schematically in Fig. 1. The overall idea is to make use of one of the
excited states of the system to induce electron transitions between the
lowest-energy states localized in different quantum dots under the influence
of resonant external perturbation (e.g., an ac electromagnetic field).
According to the laws of quantum mechanics an electron, having been "raised
up" (at some moment in time) by the perturbation from the localized state,
e.g., $|A1 \rangle$, to the excited state, e.g.,
$(|A2 \rangle + |B2 \rangle)/\sqrt{2}$, immediately becomes spread over
both dots, so that it can be subsequently "lowered down" by the same
perturbation (acting on both dots) to the localized state of the {\it other}
quantum dot, in our example $|B1 \rangle$. The physical picture of such an
effect seems quite clear, the question is only in the probability of electron
transfer between the dots.

As to the case of an {\it isolated} quantum dot, it is well known that a
periodic perturbation $\hat{F}\cos{(\Omega t)}$ with frequency
$\Omega=\epsilon_2-\epsilon_1$ (hereinafter $\hbar=1$) leads to periodic
oscillations of the probabilities $p_1(t)$ and $p_2(t)$ of detecting an
electron in levels $|1 \rangle$ and $|2 \rangle$ \cite{Landau,Flugge}. If
$p_1(0)=1$ and $p_2(0)=0$ at the initial moment, then
\begin{equation} p_2(t)=\sin^2{(\omega_R t)} ,
\label{p2t}
\end{equation}
where $\omega_R=|\langle 2|\hat{F}|1\rangle|/2$. Here
$\langle 2|\hat{F}|1\rangle$ is the matrix element of the interlevel
transition. It follows from Eq.(\ref{p2t}) that one can select the time $T$
during which the perturbation is on (for example, $T=\pi/2\omega_R$) so that
the condition $p_2(T)=1$ is satisfied (so called $\pi$-pulse). Below we shall
show that in the case of the double-dot system, the probability of electron
transfer between the localized low-lying states of quantum dots can also be
put very close to unity through the proper choice of the characteristics of
an electromagnetic pulse. Similar effect in semiconducting quantum wells has
been discussed in \cite{Kopaev}.

\vskip 6mm

\centerline{\bf 3. DESCRIPTION OF THE MODEL}

\vskip 2mm

Let the external perturbation be the classical ac electric field
${\bf E}(t)$. Then the model Hamiltonian has the form
\begin{eqnarray}
\hat{H}(t)=
\epsilon_1(\hat{a}^{+}_{A1}\hat{a}^{}_{A1}+\hat{a}^{+}_{B1}\hat{a}^{}_{B1})+
\epsilon_2(\hat{a}^{+}_{A2}\hat{a}^{}_{A2}+\hat{a}^{+}_{B2}\hat{a}^{}_{B2})-
V(\hat{a}^{+}_{A2}\hat{a}^{}_{B2}+h.c.)+
{\bf E}(t)\left[{\bf d}(\hat{a}^{+}_{A2}\hat{a}^{}_{A1}+
\hat{a}^{+}_{B2}\hat{a}^{}_{B1})+h.c.\right]~,
\label{Ham1}
\end{eqnarray}
where $\hat{a}^{+}_{\alpha i}$ $(\hat{a}^{}_{\alpha i})$ is the electron
creation (annihilation) operator for an electron in states $|\alpha i\rangle~
(\alpha = A,B;~i=1,2);~{\bf d}=\langle A2|-e{\bf r}|A1 \rangle =
\langle B2|-e{\bf r}|B1 \rangle$ is the matrix element of optical dipole
transitions $|A1\rangle \rightleftharpoons |A2\rangle$ and
$|B1\rangle \rightleftharpoons |B2\rangle$. We do not specify the spin index
explicitly since we consider a single electron whose spin projection on an
arbitrary chosen axis remains unchanged upon the action of ac electric field.
Note that in Eq.(\ref{Ham1}) we have omitted the terms describing both
tunnel and optical transitions $|A1\rangle \rightleftharpoons |B2\rangle$ and
$|B1\rangle \rightleftharpoons |A2\rangle$ since the wave functions entering
into the corresponding matrix elements are centered in different quantum dots
(one of wave functions being strongly localized within a particular dot), and
hence one can expect those matrix elements be exponentially smaller than $V$
and $d$ respectively.

Let us suppose that the external field is turned on at $t=0$ and turned off
at $t=T$, and has a frequency $\Omega$, i.e.,
\begin{equation}
{\bf E}(t)={\bf E_0}\cos{(\Omega t)}\theta(t)\theta(T-t) ,
\label{Field}
\end{equation}
where ${\bf E_0}$ is the field amplitude, $\theta(t)$ is the Heaviside step
function. The pulse duration $T$ and frequency $\Omega$ are to be derived by
maximizing the probability that an electron is transferred from the localized
low-lying state of one quantum dot to that of another dot.

It is convenient to introduce new notations for one-electron states:
\begin{equation}
|1\rangle = |A1\rangle,~~|2\rangle = |B1\rangle,~~
|3\rangle = (|A2\rangle + |B2\rangle)/\sqrt{2},~~
|4\rangle = (|A2\rangle - |B2\rangle)/\sqrt{2},~~
\label{States}
\end{equation}
and hence to replace in Eq.(\ref{Ham1}) the operators $\hat{a}^{}_{A1}$ and
$\hat{a}^{}_{B1}$ by operators $\hat{a}_1$ and $\hat{a}_2$ respectively as
well as to go from operators $\hat{a}^{}_{A2}$ and $\hat{a}^{}_{B2}$
describing the excited states of isolated quantum dots with the energy
$\epsilon_2$ to operators
$\hat{a}_3 = (\hat{a}^{}_{A2} + \hat{a}^{}_{B2})/\sqrt{2}$ and
$\hat{a}_4 = (\hat{a}^{}_{A2} - \hat{a}^{}_{B2})/\sqrt{2}$ describing the
excited sublevels of the double-dot nanostructure with energies
$\epsilon_2-V$ and $\epsilon_2+V$ respectively. Then the time-independent
part of the Hamiltonian acquires a diagonal form, while optical transitions
take place between the states $|1\rangle \rightleftharpoons |3\rangle$,
$|2\rangle \rightleftharpoons |3\rangle$,
$|1\rangle \rightleftharpoons |4\rangle$, and
$|2\rangle \rightleftharpoons |4\rangle$:
\begin{eqnarray}
\hat{H}(t)=
\epsilon_1(\hat{a}^+_1\hat{a}^{}_1+\hat{a}^+_2\hat{a}^{}_2)+
(\epsilon_2-V)\hat{a}^+_3\hat{a}^{}_3
+(\epsilon_2+V)\hat{a}^+_4\hat{a}^{}_4+
{\bf E}(t)\left[\frac{{\bf d}}{\sqrt{2}}(\hat{a}^+_3\hat{a}^{}_1+
\hat{a}^+_3\hat{a}^{}_2+\hat{a}^+_4\hat{a}^{}_1
-\hat{a}^+_4\hat{a}^{}_2)+h.c.\right].
\label{Ham2}
\end{eqnarray}

For the system under consideration, the one-electron wave function $\Psi(t)$
can be expressed at any moment as
\begin{equation}
\Psi (t) = \sum_{i=1}^{4} A_i(t) \exp(-iE_i t)|i\rangle ,
\label{Psi}
\end{equation}
where $E_i$ are eigenvalues of the stationary Schr\"odinger equation
$\hat{H}|i\rangle = E_i |i\rangle$ in the absence of an applied field
$(t\le 0)$:
\begin{equation}
E_1=\epsilon_1,~~E_2=\epsilon_1,~~E_3=\epsilon_2-V,~~E_4=\epsilon_2+V.
\label{Spectrum}
\end{equation}
The values $A_i(0)$ define the electron wave function at the initial moment;
$A_i(0)=A_i(t<0)$ since $|i\rangle$ are eigenstates of the Hamiltonian
(\ref{Ham2}) for $t\le 0$. We assume that at $t\le 0$ an electron is
localized in the ground state of the dot $A$, i.e.,
$A_1(0)=1,~A_2(0)=A_3(0)=A_4(0)=0$. The probability $p_i(t)$ to find an
electron in state $|i\rangle$ at an arbitrary time $t$ is $|A_i(t)|^2$ (it
follows from the normalization condition that $p_1(t)+p_2(t)+p_3(t)+p_4(t)=1$
at any $t$). In particular, the value of $p_2(t)$ is the probability that an
electron occupies the low-lying localized level of the dot $B$ at a time $t$.
The coefficients $A_i(t)$ in Eq.(\ref{Psi}) can be calculated by solving the
time-dependent Schr\"odinger equation
\begin{equation}
i\frac{\partial \Psi (t)}{\partial t} = \hat{H}(t)\Psi (t) ,
\label{nonstat}
\end{equation}
where $\hat{H}(t)$ is given by Eq.(\ref{Ham2}), i.e., it explicitly depends
on time at $0\le t \le T$.

\vskip 6mm

\centerline{\bf 4. RESONANT APPROXIMATION}

\vskip 2mm

In order to solve the problem analytically, we use the resonant approximation
\cite{Landau,Flugge}. We assume that the frequency $\Omega$ of ac electric
field is close to the resonant frequency
\begin{equation}
\Omega_r=\epsilon_2-V-\epsilon_1 ,
\label{frequency}
\end{equation}
i.e., to the difference between the energy $\epsilon_2-V$ of the lower
excited state $|3\rangle$ and the energy $\epsilon_1$ of the two-fold
degenerate ground state $|1\rangle$ ($|2\rangle$) so that those three states
are resonant with the ac field, while the upper excited state $|4\rangle$
with the energy $\epsilon_2+V$ is out of resonance. To be precise, the value
of $\Omega$ should be much more closer to $\epsilon_2-V-\epsilon_1$ than to
$\epsilon_2+V-\epsilon_1$, i.e., the following strong inequality should be
satisfied:
\begin{equation}
|\delta| << V  ,
\label{inequality1}
\end{equation}
where the value of
\begin{equation}
\delta = \Omega - \Omega_r
\label{delta}
\end{equation}
quantifies the offset from resonance. Since in the resonant approximation
the electron transitions $|1\rangle \rightleftharpoons |4\rangle$ and
$|2\rangle \rightleftharpoons |4\rangle$ can be ignored, the effective
Hamiltonian has the form
\begin{eqnarray}
\hat{H}(t)=
\epsilon_1(\hat{a}^+_1\hat{a}^{}_1+\hat{a}^+_2\hat{a}^{}_2)+
(\epsilon_2-V)\hat{a}^+_3\hat{a}^{}_3+
\left[\frac{\lambda}{2}\exp(-i\Omega t)(\hat{a}^+_3\hat{a}^{}_1+
\hat{a}^+_3\hat{a}^{}_2)+h.c.\right],
\label{Ham3}
\end{eqnarray}
where we have introduced the notation
\begin{equation}
\lambda = {\bf E_0 d}/\sqrt{2} .
\label{lambda}
\end{equation}

Generally speaking, the Schr\"odinger equation (\ref{nonstat}) with
Hamiltonian (\ref{Ham3}) can be reduced to the system of coupled differential
equations for the coefficients $A_i(t)$ in the expansion (\ref{Psi}) of the
wave function ($i=1-3$). It is more convenient, however, to go to a
representation with a time-independent Hamiltonian making use of the unitary
transformation
\begin{equation}
\hat{U}(t)=\exp\left[\frac{i\Omega t}{2}(\hat{a}^{+}_1\hat{a}^{}_1
+\hat{a}^{+}_2\hat{a}^{}_2-\hat{a}^{+}_3\hat{a}^{}_3)\right]
\label{transform}
\end{equation}
similar to those proposed by Galitskii {\it et al.} \cite{Galitskii} to
describe the interaction of a semiconductor with a strong electromagnetic
field. Substituting
\begin{equation}
\Psi(t)=\hat{U}(t) \tilde{\Psi} (t)
\label{Psi2}
\end{equation}
into the Schr\"odinger equation (\ref{nonstat}) for $\Psi (t)$, we obtain the
Schr\"odinger equation for $\tilde{\Psi} (t)$:
\begin{equation}
i\frac{\partial \tilde{\Psi} (t)}{\partial t} =
\hat{\tilde{H}}\tilde{\Psi} (t) ,
\label{nonstat2}
\end{equation}
with the Hamiltonian
\begin{eqnarray}
\hat{\tilde{H}}&=&\hat{U}^+(t)\hat{H}(t)\hat{U}(t)-
i\hat{U}^+(t)\frac{\partial \hat{U}(t)}{\partial t} \nonumber \\
&&=(\epsilon_1+\Omega/2)(\hat{a}^+_1\hat{a}^{}_1+\hat{a}^+_2\hat{a}^{}_2)+
(\epsilon_2-V-\Omega/2)\hat{a}^+_3\hat{a}^{}_3+
\left[\frac{\lambda}{2}(\hat{a}^+_3\hat{a}^{}_1+
\hat{a}^+_3\hat{a}^{}_2)+h.c.\right]~.
\label{Hamtilde}
\end{eqnarray}

Since the Hamiltonian $\hat{\tilde{H}}$ is independent on time, the general
solution of the time-dependent Schr\"odinger equation (\ref{nonstat2}) for
$0\le t\le T$ has the form:
\begin{equation}
\tilde{\Psi}(t) = \sum_{i=1}^{3} B_i \exp(-i\tilde{E}_it)|\tilde{i}\rangle ,
\label{Psitilde}
\end{equation}
where $|\tilde{i}\rangle$ and $\tilde{E}_i$ are the eigenstates and
eigenvalues of the stationary Schr\"odinger equation
\begin{equation}
\hat{\tilde{H}}|\tilde{i}\rangle=\tilde{E}_i|\tilde{i}\rangle~.
\label{Stationary}
\end{equation}
We seek solutions of Eq.(\ref{Stationary}) in the form
\begin{equation}
|\tilde{i}\rangle = \sum_{k=1}^{3}C_{ik}|k\rangle ,
\label{Statestilde}
\end{equation}
where $|k\rangle$ are eigenstates defined by Eq.(\ref{States}) for $t\le 0$.
Substituting Eq.(\ref{Statestilde}) into Eq.(\ref{Stationary}), we obtain a
set of equations for $C_{ik}$ and $\tilde{E}_i$:
\begin{equation}
\sum_{k=1}^{3}C_{ik}(\langle i|\hat{\tilde{H}}|k \rangle-
\delta_{ik}\tilde{E}_i)=0 ,
\label{system}
\end{equation}
where $i=1\div 3$, and $\langle i|\hat{\tilde{H}}|k \rangle$ are the matrix
elements of the Hamiltonian (\ref{Hamtilde}) in terms of the first three
basis states of (\ref{States}). The Hamiltonian matrix
$\langle i|\hat{\tilde{H}}|k \rangle$ has the form
\begin{eqnarray}
\left(
\begin{array}{cccc}
\epsilon_1+\Omega/2 &\ 0 &\ \lambda^*/2 \\
0 &\ \epsilon_1+\Omega/2 &\ \lambda^*/2 \\
\lambda/2 &\ \lambda/2 &\ \epsilon_2-V-\Omega/2 \\
\end{array}
\right)~.
\label{matrix}
\end{eqnarray}
From Eqs. (\ref{Psitilde}) and (\ref{Statestilde}) one has
\begin{equation}
\tilde{\Psi}(t) = \sum_{i=1}^{3} D_i(t) |i\rangle ,
\label{Psitilde2}
\end{equation}
where
\begin{equation}
D_i(t) = \sum_{k=1}^{3}B_kC_{ki}\exp(-i\tilde{E}_kt) .
\label{D_i}
\end{equation}

Since $\tilde{\Psi}(0)=\Psi(0)$, see Eqs. (\ref{transform}) and (\ref{Psi2}),
we have $D_i(0) = A_i(0)$, where the coefficients $A_i(0)$ determine the
state (\ref{Psi}) for $t\le 0$. Therefore, from Eq.(\ref{D_i}) we obtain the
equations that determine the coefficients $B_i$ in terms of given $A_i(0)$:
\begin{equation}
A_i(0) = \sum_{k=1}^{3}B_kC_{ki} ,
\label{A_i}
\end{equation}
whence
\begin{equation}
B_i = \sum_{k=1}^{3}A_k(0)C^{-1}_{ki} ,
\label{B_i}
\end{equation}
where
$C^{-1}$ is the matrix inverse of $C$. From Eqs. (\ref{D_i}) and (\ref{B_i})
we obtain :
\begin{equation}
D_i(t) = \sum_{k=1}^{3}\sum_{l=1}^{3}A_l(0)C^{-1}_{li}C^{}_{ki}
\exp(-i\tilde{E}_kt) .
\label{D_i2}
\end{equation}
Finally, given Eq.(\ref{Psi2}) relating the function $\tilde{\Psi}(t)$ to
$\Psi(t)$ and taking into account that the operator $\hat{U}(t)$ defined by
Eq.(\ref{transform}) is unitary, we obtain an expression for the probability
$p_i(t)$ for the transition to the state $|i\rangle$:
\begin{equation}
p_i(t)=|D_i(t)|^2 .
\label{probability}
\end{equation}

\vskip 6mm

\centerline{\bf 5. RESULTS AND DISCUSSION}

\vskip 2mm

Since we assume (see above) that at $t \le 0$ an electron is localized in the
level $|1\rangle$, the lowest energy level of the dot $A$, i.e, $A_1(0)=1$
and $A_2(0)=A_3(0)=0$, expressions (\ref{D_i2}) and (\ref{probability})
simplify somewhat. Having calculated the eigenvalues $\tilde{E}_i$ and the
matrix of eigenvectors $C_{ik}$ from Eq.(\ref{system}), we obtain from
Eqs. (\ref{D_i2}) and (\ref{probability}) the expressions for the
probabilities of transitions from the state $\Psi(0)=|1\rangle$ to the state
$|i\rangle$:
\begin{eqnarray}
&&p_1(t)=\cos^4(\omega^{}_R t)-
\sin^2(\frac{\delta t}{4})\cos(2\omega^{}_R t)+
\frac{\delta^2}{64\omega_R^2}\sin^2(2\omega^{}_R t)+
\frac{\delta}{8\omega^{}_R}\sin(\frac{\delta t}{2})\sin(2\omega^{}_R t) ,
\nonumber \\
&&p_2(t)=\sin^4(\omega^{}_R t)+
\sin^2(\frac{\delta t}{4})\cos(2\omega^{}_R t)+
\frac{\delta^2}{64\omega_R^2}\sin^2(2\omega^{}_R t)-
\frac{\delta}{8\omega^{}_R}\sin(\frac{\delta t}{2})\sin(2\omega^{}_R t) ,
\nonumber \\
&&p_3(t)=\frac{1}{2}\left(1-\frac{\delta^2}{16\omega^{}_R}\right)
\sin^2(2\omega^{}_R t) ,
\label{probability2}
\end{eqnarray}
where $\delta$ is defined by Eqs. (\ref{frequency}) and (\ref{delta}), and
\begin{equation}
\omega^{}_R=\frac{\sqrt{\delta^2+2|\lambda|^2}}{4} .
\label{omega_R}
\end{equation}
In a particular case of exact resonance ($\delta=0$), one has from
Eq.(\ref{probability2}):
\begin{eqnarray}
&&p_1(t)=\cos^4(\omega^{}_R t) , \nonumber \\
&&p_2(t)=\sin^4(\omega^{}_R t) , \nonumber \\
&&p_3(t)=\frac{1}{2}\sin^2(2\omega^{}_R t) .
\label{probability3}
\end{eqnarray}

We are interested mainly in the probability $p_2(t)$ of electron transfer to
the level $|2\rangle$, the lowest energy level of the dot $B$. It follows
from Eq.(\ref{probability3}) that $p_2(t)=1$ at $t=T_n$, where
\begin{equation}
T_n=\frac{\pi}{2\omega^{}_R} + \frac{\pi n}{\omega^{}_R} ,
\label{Time}
\end{equation}
and $n$ is an integer. Hence, after the applied field is off at $t=T_n$, the
electron will stay in the state $|2\rangle$ since this state is the
eigenstate of the Hamiltonian (\ref{Ham2}) in the absence of external
perturbation. So, if a characteristic time of electron tunneling between the
states $|1\rangle$ and $|2\rangle$, the ground states of the dots $A$ and $B$
respectively, is long enough (i.e., if the two dots are placed far apart from
each other and/or are separated by relatively high energy barrier), the
electron remains, in fact, localized in the dot $B$.

When the frequency is offset from resonance ($\delta \neq 0$), the value of
$p_2(T_n)$ derived from Eq.(\ref{probability2}) deviates from unity by a
quantity of order $\delta^2 T_n^2$. At a given value of $\delta$, the
probability $p_2(T_n)$ has a maximum for $n=0$, see Eq.(\ref{Time}), i.e.,
for the perturbation duration time $T_0=\pi/2\omega^{}_R$:
\begin{equation}
p_2(T_0)=1-\frac{\pi^2}{64}\frac{\delta^2}{\omega^{2}_R} .
\label{p_2}
\end{equation}
Taking Eq.(\ref{omega_R}) into account, we are led to the following
inequality
\begin{equation}
|\delta| << |\lambda|
\label{inequality2}
\end{equation}
which should hold in order that the probability $p_2(T_0)$, Eq.(\ref{p_2}),
be very close to unity.

Note that the perturbing ac field acting for a {\it finite} period of time,
$T_0$, contains harmonics in the frequency range $\delta\omega\approx 1/T_0$.
In order that the approximating Hamiltonian, Eq.(\ref{Ham3}), be valid, the
bandwidth $\delta\omega$ should be much smaller than the interval $2V$
between the energies of excited states $|3\rangle$ and $|4\rangle$ since
otherwise the external field will mix all the states $|1\rangle$,
$|2\rangle$, $|3\rangle$, and $|4\rangle$, see Eq.(\ref{States}). Besides,
the time $T_0$ needed to transfer an electron from the ground state
$|1\rangle$ of the dot $A$ to the ground state $|2\rangle$ of the dot $B$
should be much shorter than the lifetime $\tau^*$ of electron in the
"auxiliary" excited state $|3\rangle$ of the nanostructure, see Sec.2, since
otherwise the probability of photon/phonon emission at $t \ll T_0$ is high,
resulting in decoherence and breakdown of unitary electron evolution under
the influence of ac field. Hence, taking into account that
$T_0 \sim 1/|\lambda|$, see Eqs. (\ref{omega_R}), (\ref{Time}), and
(\ref{inequality2}), we arrive at the following inequalities:
\begin{equation}
\frac{1}{\tau^*} << |\lambda| << V .
\label{inequality3}
\end{equation}
Finally, combining Eqs. (\ref{inequality1}), (\ref{inequality2}), and
(\ref{inequality3}) together, one has the following conditions for ({\it i})
applicability of the resonant approximation to the description of
electronic transitions in the nanostructure under consideration and
({\it ii}) proximity of the probability of electron transfer from one dot
to another to unity:
\begin{equation}
\frac{1}{\tau^*}, \delta << |\lambda| << V .
\label{inequality4}
\end{equation}

We note that conditions (\ref{inequality4}) imposed on the frequency,
duration and amplitude of electromagnetic pulse can be fulfilled in the
experiment. Indeed, if $V\sim 10^{-3}$ eV (see estimates in Sec.2), we should
have, e.g., $|\lambda|\approx 10^{-5}-10^{-4}$ eV and
$|\delta|\approx 10^{-7}-10^{-6}$ eV (our numerical calculations have shown
that at $|\lambda|/V=0.1$ and $|\delta|/|\lambda|=0.1$ the value of
$p_2(T_0)$ is about 0.99). Since the parameter $|\lambda|$ is of the order of
a product of the electric field amplitude $E_0$ and the optical dipole matrix
element $d$, see Eq.(\ref{lambda}), and $d\approx ea$, for a characteristic
quantum dot size $a\approx 10$ nm we obtain $E_0\approx 10-100$ V/cm, which
can be easily realized experimentally. As to the condition imposed on
$\delta$, for assumed value of $V$ one finds that the frequency $\Omega$ of
the external source should be accurate to within $10^9$ s$^{-1}$ or better.
Such an accuracy can be obtained by modern experimental methods. Besides, the
resonance condition can probably be achieved by varying the energy difference
between the size-quantized levels via applying a gate voltage to the
nanostructure.

From symmetry considerations it is clear that if at $t\leq 0$ an electron is
in the state $|2\rangle$ (i.e. in the lowest-energy state of the dot $B$), it
will take the same time $T_0$ to transfer it to the state $|1\rangle$ (i.e.
to the lowest-energy state of the dot $A$) as the time needed for electron
transfer from the state $|1\rangle$ to the state $|2\rangle$, see above.
Hence, if initially an electron is in an arbitrary superposition of states
$|1\rangle$ and $|2\rangle$, i.e. $\Psi(0)=\alpha|1\rangle+\beta|2\rangle$
where $\alpha$ and $\beta$ are complex numbers such that
$|\alpha|^2+|\beta|^2=1$, then $\Psi(T_0)=\beta|1\rangle+\alpha|2\rangle$.

Now it is in order to mention a possible application of the effect of ac
field-induced electron transfer between two quantum dots to the so called
"quantum computation" \cite{Lloyd,Bennett}. Indeed, if the states
$|1\rangle$ and $|2\rangle$, i.e. electron locations in dots $A$ and $B$, are
viewed as the Boolean states 0 and 1 respectively, then their linear
combination $\alpha|1\rangle+\beta|2\rangle$ may be viewed as a "quantum bit"
("qubit"). In its turn, an action of the resonant (in the sense discussed
above) ac field on the double-dot nanostructure may be considered as a
unitary operation NOT over the qubit:
\begin{eqnarray}
U_{NOT}
\left(
\begin{array}{cccc}
\alpha \\
\beta \\
\end{array}
\right)=
\left(
\begin{array}{cccc}
\beta \\
\alpha \\
\end{array}
\right).
\label{matrix1}
\end{eqnarray}
Such an operation is non-dissipative (reversible). Hence, the double-dot
nanostructure can function as a reversible logic gate NOT (inverter), in
contrast to a number of dissipative (non-reversible) logic circuits proposed
in the literature (see, e.g., \cite{Nomoto,Bandyo1,Molotkov,Openov}).

Various schemes for realizing reversible quantum logic gates have also been
discussed (see, e.g., \cite{Lloyd,DiVincenzo1,Cirac,Kane,DiVincenzo2}) and
demonstrated experimentally \cite{Turchette,Monroe,Chuang}. Almost all of
these schemes are based on encoding qubits in either photon states or in
nuclear spins. From the perspective of high density computational circuits,
the reversible logic gates based on single electrons in quantum dots seem to
be very appealing. Several quantum gate mechanisms based on electron spins in
{\it strongly} coupled adjacent quantum dots have been proposed (see, e.g.,
\cite{Bandyo2,Bychkov,Loss,Burkard}). In our opinion, encoding qubits in
electron locations (i.e., in fact, in ground states of {\it weakly} coupled
well separated quantum dots) rather than in electron spins may have an
advantage that such qubits are well defined and can be expected to have long
dephasing times. Besides, the measurement procedure may appear to be more
straightforward than in the case of spin-based qubits.

Of course, a quantum inverter is the simplest logic gate. It operates with
one qubit only and is not a {\it universal} gate, i.e. it cannot be
considered as the fundamental building block of a quantum computer. However
the ideas presented in this paper can be used to construct more complex gates
consisting of several quantum dots and operating with more than one qubit,
e.g. XOR (controlled-NOT) gate.

In conclusion, we have shown that a resonant electron transfer between the
states localized in distant quantum dots can take place upon the influence of
a resonant ac field with properly chosen characteristics (amplitude,
frequency, and duration). Such a transfer occurs via an excited bound state
of the double-dot nanostructure delocalized over both dots. Although only the
simplest case that each dot has two size-quantized energy levels has ben
considered, it is clear that, in general, the resonant electron transfer
between the dots can be assisted by any excited level whose energy lies close
to the top of the barrier separating the dots if the appropriate resonance
conditions are satisfied. Since an electron can be in a quantum-mechanical
superposition of states localized in different dots, the double-dot
nanostructure can play a role of a reversible logic gate NOT operating the
quantum bits.

\vskip 6mm

\centerline{\bf ACKNOWLEDGMENTS}

\vskip 2mm

This work was supported by the Russian Federal Program "Integration", the
Russian Foundation for Fundamental Research under Grant No 96-02-18918 and by
the Russian State Program "Advanced Technologies and Devices in Micro- and
Nanoelectronics". The author would like to thank V. V. V'yurkov,
V. F. Elesin, and S. N. Molotkov for useful discussions.

\newpage
\centerline{\bf FIGURE CAPTIONS}
\vskip 2mm

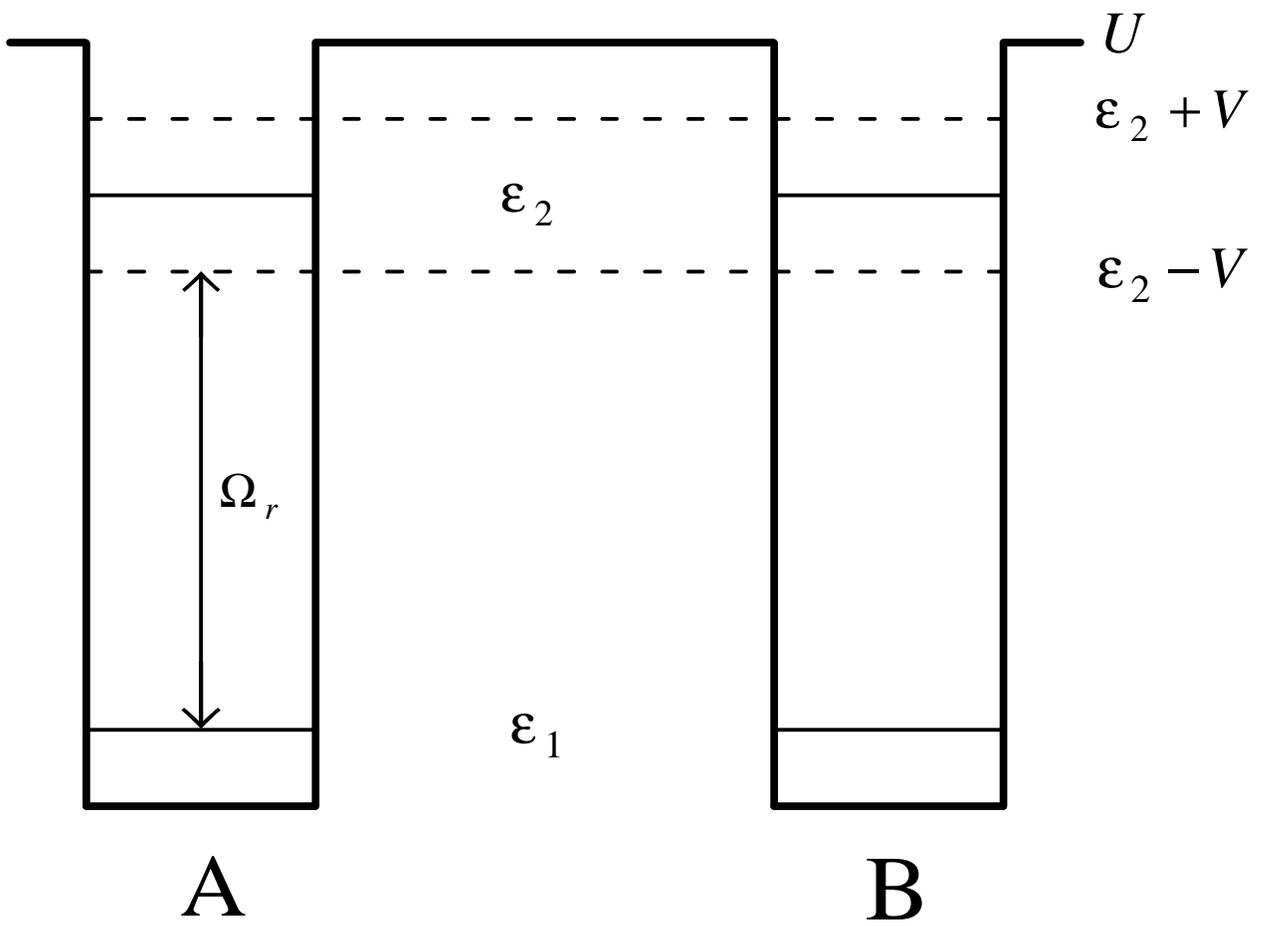
Fig.1. Energy levels diagram of a nanostructure composed of two identical
quantum dots, $A$ and $B$, see text for details.


\vskip 6mm

\begin{references}

\bibitem{Luth} H. L\"uth, Phys. Stat. Sol. (b) {\bf 192}, 287 (1995).
\bibitem{Krash} A. V. Krasheninnikov, S. N. Molotkov, S. S. Nazin, and
L. A. Openov, Zh. Eksp. Teor. Fiz. {\bf 112}, 1257 (1997) [JETP {\bf 85},
682 (1997)].
\bibitem{Grossmann} F. Grossmann, T. Dittrich, P. Jung, and P.H\"anggi,
Phys. Rev. Lett. {\bf 67}, 516 (1991).
\bibitem{Landau} L. D. Landau and E. M. Lifshitz, 3rd ed., Nauka, Moscow
(1974) [Pergamon Press, New York (1977)].
\bibitem{Nomoto} K. Nomoto, R. Ugajin, T. Suzuki, and I. Hase, J. Appl. Phys.
{\bf 79}, 291 (1996).
\bibitem{Kopaev} A. A. Gorbatsevich, V. V. Kapaev, and Yu. V. Kopaev,
Zh. Eksp. Teor. Fiz. {\bf 107}, 1320 (1995) [JETP {\bf 80}, 734 (1995)].
\bibitem{Flugge} S. Fl\"ugge, {\it Practical Quantum Mechanics II},
Springer-Verlag, Berlin-Heidelberg-New York (1971).
\bibitem{Galitskii} V. M. Galitskii, S. P. Goreslavskii, and V. F. Elesin,
Zh. Eksp. Teor. Fiz. {\bf 57}, 207 (1969) [JETP {\bf 30}, 117 (1970)].
\bibitem{Lloyd} S. Lloyd, Science {\bf 261}, 1569 (1993).
\bibitem{Bennett} C. H. Bennett, Physics Today, October 1995, p.24.
\bibitem{Bandyo1} S. Bandyopadhyay, V. P. Roychowdhury, and X. Wang,
Phys. Low-Dim. Struct. {\bf 8/9}, 28 (1995).
\bibitem{Molotkov} S. N. Molotkov and S. S. Nazin, Pis'ma v ZhETF {\bf 62},
256 (1995) [JETP Lett. {\bf 62}, 273 (1995)]; ZhETF {\bf 110}, 1439 (1996)
[JETP {\bf 83}, 794 (1996)].
\bibitem{Openov} A. V. Krasheninnikov and L. A. Openov, Pis'ma v ZhETF
{\bf 64}, 214 (1996) [JETP Lett. {\bf 64}, 231 (1996)].
\bibitem{DiVincenzo1} D. P. DiVincenzo, Science {\bf 270}, 255 (1995).
\bibitem{Cirac} J. I. Cirac and P. Zoller, Phys. Rev. Lett. {\bf 74}, 4091
(1995).
\bibitem{Kane} B. E. Kane, Nature {\bf 393}, 133 (1998).
\bibitem{DiVincenzo2} D. P. DiVincenzo, Nature {\bf 393}, 113 (1998).
\bibitem{Turchette} Q. A. Turchette, C. J. Hood, W. Lange, H. Mabuchi, and
H. J. Kimble, Phys. Rev. Lett. {\bf 75}, 4710 (1995).
\bibitem{Monroe} C. Monroe, D. M. Meekhof, B. E. King, W. M. Itano, and
D. J. Wineland, Phys. Rev. Lett. {\bf 75}, 4714 (1995).
\bibitem{Chuang} I. L. Chuang, L. M. K. Vandersypen, X. Zhou, D. W. Leung,
and S. Lloyd, Nature {\bf 393}, 143 (1998).
\bibitem{Bandyo2} S. Bandyopadhyay and V. P. Roychowdhury,
Superlattices and Microstructures {\bf 22}, 411 (1997).
\bibitem{Bychkov} A. M. Bychkov, L. A. Openov, and I. A. Semenihin,
Pis'ma v ZhETF {\bf 66}, 275 (1997) [JETP Lett. {\bf 66}, 298 (1997)].
\bibitem{Loss} D. Loss and D. P. DiVincenzo, Phys. Rev. A {\bf 57}, 120
(1998).
\bibitem{Burkard} G. Burkard, D. Loss, and D. P. DiVincenzo, Phys. Rev. B
{\bf 59}, 2070 (1999).

\end{references}
\end{document}